\documentclass[11pt]{article}
\usepackage{amsfonts}
\usepackage{amssymb}
\usepackage{amsmath}
\usepackage{latexsym}
\def \Z {\mathbb{Z}}
\def \R {\mathbb{R}}

\def \la {\lambda}

\newcommand{\bequ}{\begin{equation}}
\newcommand{\eequ}{\end{equation}}
\newcommand{\barr}{\begin{array}}
\newcommand{\earr}{\end{array}}
\newcommand{\bea}{\begin {eqnarray}}
\newcommand{\eea}{\end {eqnarray}}

\begin{document}

\title{Non-equilibrium Statistical Mechanics of Anharmonic Crystals with Self-consistent Stochastic Reservoirs}
\author{Emmanuel Pereira$^{1}$ and Ricardo Falcao$^{2}$\\
{\small Departamento de F\'{\i}sica-ICEx, UFMG, CP 702, 30.161-970 Belo Horizonte MG, Brazil}\\
{\small e-mail: $^{1}$ emmanuel@fisica.ufmg.br;
$^{2}$rfalcao@fisica.ufmg.br}} \maketitle

\begin{abstract}

We consider a $d$-dimensional crystal with an arbitrary harmonic
interaction and an anharmonic on-site potential, with stochastic
Langevin heat bath at each site. We develop an integral formalism
for the correlation functions that is suitable for the study of
their  relaxation (time decay) as well as their behavior in space.
Furthermore, in a perturbative analysis, for the one-dimensional
system with weak coupling between the sites and small quartic
anharmonicity, we investigate the steady state and show that the
Fourier's law holds. We also obtain an expression for the thermal
conductivity (for arbitrary next-neighbor interactions) and give
the temperature profile in the steady state.

\vspace{.5cm}


\vspace{.5cm}

\noindent {\bf Short title:} Anharmonic Crystals with
Self-consistent Reservoirs
\end{abstract}

\section{Introduction}

We are surrounded by phenomena involving non-equilibrium
processes, but our understanding of such systems, i.e., the number
of models that permit detailed calculations, is very limited. In
particular, it is unknown a simple way of finding the properties
in the steady states: e.g., a rigorous derivation of the
(phenomenological) Fourier's law from a microscopic anharmonic
Hamiltonian model has not been established up to now (see
\cite{BLR}, \cite{LLP} for a review). It makes the analysis of
simple dynamical models describing non-equilibrium processes a
problem of interest.

A commonly studied microscopic model is the Hamiltonian chain (or
its d-dimensional version) of N interacting oscillators coupled to
heat baths at each site or at the boundaries only, and its
anharmonic version with small quartic on-site interactions.

For the harmonic case of the model with thermal reservoirs at the
boundaries, the covariance of the stationary state has been
calculated in \cite{RLL}, long time ago. There, it is shown that
the heat current is independent on the length of the chain, and
so, the Fourier's law does not hold. The temperature profile is
also computed in \cite{RLL}: the temperature is essencially
constant in the interior of the chain, but decreases exponentially
close to the hotter bath and increases close to the opposite end.
I.e., the profile has the lowest temperature near the hottest
reservoir and the highest temperature near the coldest reservoir.
For the anharmonic case, there are interesting and recent results.
The existence of steady states is proved in \cite{EPRCMP}, and the
positivity of entropy production in \cite{EPRJSP}. Numerical
results strongly suggest that the Fourier's law holds in such a
case \cite{HuPRE}, \cite{AKAnPh}, but,  in the opposite direction,
 a perturbative analysis \cite{LSJSP} shows that the heat
current does not depend on the size of the system. Also in the
perturbative study \cite{LSJSP}, as in harmonic case \cite{RLL},
the temperature profile (discarding the exponential decay in the
bulk of the chain) is in the ``wrong'' way: the hottest
temperature is near the coldest bath, and vice-versa. In short, it
is unclear whether the Fourier's law holds or not in such
anharmonic models. It is worth to recall that other results also
indicate that it is wrong the  opinion that the sole anharmonicity
of the on-site potential shall ensure normal heat conductivity in
some commonly used models \cite{SG}.

The harmonic crystal model with next-neighbor interactions and
heat bath at each site has been recently analyzed in \cite{BLL}.
It is proved, for a uniquely fixed temperature profile leading to
the steady state (given the temperatures at the boundaries),  that
the heat current satisfies the Fourier's law. For the case of more
intricate interactions (intense and beyond  next-neighbor sites),
for a chain with some few sites, some results  presented in
\cite{EZ} indicate that there is a ``strange'' heat flux in the
harmonic network (and the authors claim that the results persist
under weak anharmonic perturbations): inside the chain, the
direction of the heat fluxes can not (in general) be supposed from
the temperature of the heat baths.

In the present paper, also with the aim of studying the dynamics
of simple microscopic models in order to understand properties of
non-equilibrium systems, we study the anharmonic version of this
crystal with stochastic Langevin heat bath at each site (model
named as crystal with self-consistent reservoirs). We describe an
approach and obtain an integral formalism suitable for the study
of the correlation functions (of the d-dimensional system with
quite general interactions). Furthermore, using perturbative
calculations, for a weak coupling between the sites and a weak
anharmonic potential, we show (for the one-dimensional system)
that the Fourier's law still holds. That is, we show (at least up
to first order in the perturbative computation) that the Fourier's
law is valid for this microscopic anharmonic Hamiltonian model. We
also obtain an expression for the thermal conductivity (for
next-neighbor interactions which may arbitrarily change along the
chain), and give the temperature profile in the steady state. For
the simpler case of next-neighbor interactions constant along the
chain, our results (considering the anharmonic model) coincide
with those of the harmonic case recently described in \cite{BLL}.

The rest of the paper is organized as follows. In section 2 we
present the model and some expressions for the energy current. The
integral formalism for the correlation functions is developed  in
section 3. In section 4, in a perturbative computation, we analyze
the energy current in the steady state and the Fourier's law. In
section 5 we argue on the reliability of the perturbative results
and present some concluding remarks.

\section{The Model and Initial Considerations}

Lets us introduce the model to be analyzed here and some
expressions for the energy current. We consider the stochastic
Langevin dynamics of an anharmonic crystal, i.e. a scalar field
lattice model with unbounded spin variables in a d-dimensional
lattice space box $\Lambda \subset \Z^d$, with stochastic heat
bath at each site. Precisely, we take a system of $N$ oscillators
with Hamiltonian
\begin{equation}
H(q,p)=\sum_{j=1}^{N}\frac{1}{2}\left [p_j^2+Mq_j^2\right ] +
\frac{1}{2}\sum_{j\neq l=1}^{N}q_lJ_{lj}q_j+\sum_{j=1}^{N}\lambda
\mathcal{P}(q_j) , \label{Hamiltonian}
\end{equation}
where $M>0$ , $\mathcal{P}$ gives the anharmonic on-site
perturbation (e.g., $\mathcal{P}(q_j)=q_j^4$), and we consider the
time evolution given by the stochastic differential equations
\begin{eqnarray}
dq_j&=&p_jdt , \quad\quad\quad j=1,\ldots,N, \label{eqdynamics}\\
dp_j&=&-\frac{\partial H}{\partial q_j}dt-\zeta
p_jdt+\gamma^{1/2}_jdB_j  , \quad\quad\quad j=1,\ldots,N,
\nonumber
\end{eqnarray}
where $B_j$ are independent Wiener processes, i.e., $dB_j/dt$ are
independent white noises; $\zeta$ is the heat bath coupling; and
$\gamma_j=2\zeta T_j$, where $T_j$ is the temperature of the j-th
heat bath.

To describe the energy current in the system, we write the local
energy of the spin (oscillator) $j$ as
\begin{equation}
H_j(q,p)=\frac{1}{2}p_j^2+U^{(1)}(q_j)+\frac{1}{2}\sum_{l\neq
j}U^{(2)}(q_j-q_l) , \label{lenergy}
\end{equation}
where the expression for $U^{(1)}$ and $U^{(2)}$ follow
immediately from (\ref{Hamiltonian}) and $\sum_{j=1}^{N}H_j=H$.
Then, we have
\begin{equation}
\left < \frac{dH_j(t)}{dt} \right > =\left < R_j(t) \right > -
\left <\mathcal{F}_{j<}-\mathcal{F}_{j>} \right >, \label{venergy}
\end{equation}
where $\left <\cdot\right>$ denotes the expectation with respect
to the noise distribution, and
\begin{equation}
\left <R_j(t)\right >=\zeta\left ( T_j- \left <p_j^2\right
> \right)\label{reser}
\end{equation}
gives the energy flux from the j-th reservoir to the j-th site.
The energy current inside the system is given by $\mathcal{F}_j$,
where
\begin{eqnarray}
\mathcal{F}_{j<}&=&\sum_{l>j}\nabla U^{(2)}\left ( q_j-q_l\right
)\frac{p_j+p_l}{2}, \label{flux}\\
\mathcal{F}_{j>}&=&\sum_{l<j}\nabla U^{(2)}\left ( q_l-q_j\right
)\frac{p_l+p_j}{2}. \nonumber
\end{eqnarray}
In particular, in the steady state we have $\left < dH_j(t)/dt
\right >=0$. We will turn to these expressions to discuss the
Fourier's law later.

\section{The Integral Formalism for the Correlation Functions}

For convenience, we introduce the phase-space vector $\phi=(q,p)$
with $2N$ coordinates and write the equation for the dynamics (2)
as
\begin{equation}
\dot{\phi}=-A\phi-\la \mathcal{P}'(\phi)+\sigma\eta,
\label{dynamics}
\end{equation}
where $A=(A^0+\mathcal{J})$ and $\sigma$ are $2N\times 2N$
matrices given by
\begin{eqnarray}
A^0=\left (
\begin{array}{cc}
0 & -I \\
\mathcal{M} & \Gamma \end{array} \right ), & \mathcal{J}=\left (
\begin{array}{cc}
0 & 0 \\
J & 0
\end{array}\right ), & \sigma=\left (
\begin{array}{cc}
0 & 0 \\
0 & \sqrt{2\Gamma\mathcal{T}}
\end{array}\right ).\label{defin}
\end{eqnarray}
$I$ above is the unit $N\times N$ matrix; $J$ is the $N \times N$
matrix for the two site interaction $J_{lj}$
(see(\ref{Hamiltonian})); and $\mathcal{M}$,$\Gamma$,$\mathcal{T}$
are diagonal $N \times N$ matrices:
$\mathcal{M}_{jl}=M\delta_{jl}$ , $\Gamma_{jl}=\zeta\delta_{jl}$ ,
$\mathcal{T}_{jl}=T_j\delta_{jl}$. $\eta$ are independent
white-noises; $\mathcal{P}'(\phi)$ is a $2N \times 1$ matrix with
$\mathcal{P}'(\phi)_j=0$ for $ j=1,\ldots,N$ and
\begin{equation}
\mathcal{P}'(\phi)_i=\frac{d\mathcal{P}(\phi_{i-N})}{d\phi_{i-N}}
\quad  for  \quad i=N+1,\ldots,2N.
\end{equation}

To describe the dynamics we first consider the system without the
coupling $J$ among the sites and without the anharmonic
perturbation $(\lambda=0)$ (interactions which we include in a
second step). Then the (straightforward) solution of
(\ref{dynamics}) above with $J\equiv0$, $\lambda=0$ is the
Ornstein-Uhlenbeck process given by
\begin{equation}
\phi(t)=e^{-tA^0}\phi(0)+\int_{0}^{t}ds e^{-(t-s)A^0}\sigma\eta(s)
. \label{Orns}
\end{equation}
For simplicity we take $\phi(0)=0$. The covariance of this
Gaussian process evolves as
\begin{eqnarray}
\left < \phi(t)\phi(s)\right >_{0} \equiv \mathcal{C}(t,s)=\left
\{
\begin{array}{c}
e^{-(t-s)A^0}\mathcal{C}(s,s) \quad\quad\quad t\geq s, \\
\mathcal{C}(t,t)e^{-(s-t)A^{0^{T}}} \quad\quad\quad t\leq s,
\end{array} \right . \label{covariance}
\end{eqnarray}
\begin{equation}
\mathcal{C}(t,t)=\int_{0}^{t}ds e^{-sA^0}\sigma e^{-sA^{0^T}}.
\end{equation}
It is easy to see (e.g. diagonalizing $A^0$ ) that
\begin{eqnarray}
\exp\left (-tA^0 \right )=e^{-t\frac{\zeta}{2}}\cosh(t\rho)\left
\{ I +\frac{\tanh(t\rho)}{\rho}\left (
\begin{array}{cc}
\frac{\zeta}{2}I & -I \\
-\mathcal{M} & -\frac{\zeta}{2}I
\end{array} \right ) \right \},
\end{eqnarray}
(and a similar expression follows for the transpose
$\exp(-tA^{0^{T}})$ ), where $I$ is the $N\times N$ unit matrix,
etc; $\rho=\left ( (\zeta/2)^2-M\right )^{1/2}$ (we assume that
$\left (\zeta/2\right )^2>M>0$). In this simple case (of
$J\equiv0$, $\lambda=0$), as $t\rightarrow\infty$ we have a
convergence to equilibrium (any single site is isolate) and the
stationary state is Gaussian, with mean zero and covariance
\begin{eqnarray}
C=\int_{0}^{\infty}ds ~~~~ e^{-sA^0}\sigma e^{-sA^{0^T}}=\left (
\begin{array}{cc}
\frac{\mathcal{T}}{M} & 0 \\
0 & \mathcal{T}
\end{array} \right ),\label{covariances}
\end{eqnarray}
where, again, $\mathcal{T}$ is a diagonal matrix with elements
$T_i\delta_{ij}$ (in short, for any site we have a Gibbs measure
at temperature $T_i$).

To introduce the coupling interactions and the anharmonic
potential, we use a tool of general theory of stochastic
differential equations, namely the Girsanov theorem \cite{O}. It
gives a measure $\rho$ for the new process (\ref{dynamics}) as a
``perturbation" of the measure $\mu_{\mathcal{C}}$ associated to
the process with $J\equiv 0$ , $\lambda=0$.  Precisely, for any
measurable set $A$, it states that $\rho(A)=E_0(1_{A}Z(t))$, where
$E_{0}$ is the expectation for $\mu_{\mathcal{C}}$ (the process
with $J\equiv 0$ , $\lambda=0$); $1_{A}$ denotes the
characteristic function, and
\begin{eqnarray}
Z(t)&=&\exp\left(\int_{0}^{t}u\cdot dB
-\frac{1}{2}\int_{0}^{t}u^2ds\right), \label{girs}\\
\gamma_i^{1/2}u_i&=&-\mathcal{J}_
{ik}\phi_k-\lambda\mathcal{P}'(\phi)_i ,\nonumber
\end{eqnarray}
(the inner products above are in $\R^{2N}$). From (\ref{defin})
and the expression above for $u_i$, note that $u_i$ is
nonvanishing only for $i>N$ (i.e., $i\in [N+1,N+2,\ldots,2N ]$).
In what follows we will use the following index notation: $i$ for
index values in the set $[N+1,N+2,\ldots,2N]$, $j$ for  values in
the set $[1,2,\ldots,N]$, and $k$ for values in $[1,2,\ldots,2N]$.

For clearness, let us rewrite the stochastic equations for the
initial process (with $J\equiv 0$ , $\lambda=0$) as
\begin{eqnarray}
d\phi_j&=&-A^0_{jk}\phi_kdt, \quad j\in [1,\ldots,N], \label{stoeq}\\
d\phi_i&=&-A^0_{ik}\phi_kdt+\gamma_i^{1/2}dB_i, \quad
i\in[N+1,\ldots,2N], \nonumber
\end{eqnarray}
the sum over $k$ (in $[1,2\ldots,2N])$ is assumed above (as well
as obvious sum over some indices in what follows).

Turning to the terms in $Z(t)$ we have
\begin{eqnarray*}
u_idB_i=\gamma_i^{-1/2}u_i\cdot\gamma_i^{1/2}dB_i&=&\gamma_i^{-1/2}u_i\cdot\left
(d\phi_i+A^0_{ik}\phi_kdt \right )\\
&=&\left
(-\gamma_i^{-1}\mathcal{J}_{ij}\phi_j-\gamma_i^{-1}\lambda\mathcal{P}'(\phi)_i\right
) \left (d\phi_i+A^0_{ik}\phi_kdt\right ) ,
\end{eqnarray*}
that follows from (\ref{girs}) and (\ref{stoeq}) above. We still
use It\^o formula to write the terms with $d\phi_i$ as
\begin{eqnarray*}
-\gamma_i^{-1}\mathcal{J}_{ij}\phi_jd\phi_i&=&-dF_1-\gamma_i^{-1}\phi_i\mathcal{J}_{ij}A^0_{jk}\phi_kdt,
\\
F_1(\phi)&=&\gamma_k^{-1}\phi_i\mathcal{J}_{ij}\phi_j .
\end{eqnarray*}
With similar manipulations we obtain
\begin{eqnarray}
\lefteqn{Z(t)\equiv \exp\left(\int_{0}^{t}u\cdot
dB-\frac{1}{2}\int_{0}^{t}u^2ds\right)} \nonumber \\
 & = & \exp\left\{ -F_1(\phi(t))+F_1(\phi(0))-\lambda
F_2(\phi(t))+\lambda
F_2(\phi(0))\right\}\nonumber \\
 & & \times \exp\left\{-\int_{0}^{t}W_J(\phi(s))ds - \int_{0}^{t}\lambda
W_{\lambda}(\phi(s))ds - \int_{0}^{t}\lambda W_{\lambda
J}(\phi(s))ds \right \} ,
\end{eqnarray}
with
\begin{eqnarray*}
F_1(\phi(t))&=&\gamma_i^{-1}\phi_i(t)\mathcal{J}_{ij}\phi_j(t) ,\quad F_2(\phi(t))=\gamma_i^{-1}\mathcal{P}'(\phi)_i(t)\phi_i(t) \\
W_J(\phi(s))&=&\gamma_i^{-1}\phi_i(s)\mathcal{J}_{ij}A^0_{jk}\phi_k(s)+\phi_k(s)A^{0^T}_{ki}\gamma_i^{-1}\mathcal{J}_{ij}\phi_j(s)
+ \\
&&+\frac{1}{2}\phi_{j'}(s)\mathcal{J}_{j'i}^{T}\gamma_i^{-1}\mathcal{J}_{ij}\phi_j(s), \\
\lambda W_{\lambda}(\phi(s))&= & \lambda \gamma_{i}^{-1} \phi_i(s)
\mathcal{P}'' (\phi)_i(s) A^0_{i-N,k} \phi_k(s)+ \lambda
\gamma_{i}^{-1} \mathcal{P}' (\phi)_i(s) A^0_{ik} \phi_k(s)+
\\
&&+\frac{1}{2}\lambda^2\gamma_i^{-1}(\mathcal{P}'(\phi)_i)^2(s)A^0_{i-N,k}\phi_k(s).
\\
\lambda W_{\lambda J} (\phi(s)) & = & \lambda \gamma_{i}^{-1}
\mathcal{P}' (\phi)_i(s) \mathcal{J}_{ij}\phi_j(s).
\end{eqnarray*}
And so, for the expectations, considering the process  with
coupling between sites and anharmonic perturbation, we have, e.g.
for the two-point function,
\begin{equation}
\left < \phi_u(t_1)\phi_v(t_2) \right
>=\int\phi_u(t_1)\phi_v(t_2)Z(t)d\mu_{\mathcal{C}}(\phi) , ~~~~
t_{1},t_{2}<t.
\end{equation}

The formula above, a Feynman-Kac type integral representation, is
suitable for the study of general n-point correlation functions:
for the analysis of their time decay (relaxation properties),
space behavior, etc. In particular, we will analyze the energy
current in the steady state, problem that involves the
investigation of terms such as $\lim_{t\rightarrow\infty}\left
<\phi_i(t)\phi_j(t)\right
>$, see (\ref{flux}) .

\section{The Heat Flow and Fourier's Law}

To study the heat flow in the steady state we need to analyze the
two-point correlation functions given by formula (\ref{flux}). The
averages over the stationary distributions will be obtained as the
limit
\begin{equation*}
\left <\phi_u\phi_v\right>=\lim_{t\rightarrow\infty}\left
<\phi_u(t)\phi_v(t)\right
>=\lim_{t\rightarrow\infty}\int\phi_u(t)\phi_v(t)Z(t)d\mu_{\mathcal{C}}(\phi).
\end{equation*}
We will establish conditions for the convergence to the steady
state later.

To carry out the computation, note that $\mathcal{C}(t,s)$, given
by (\ref{covariance})-(\ref{covariances}), may be written as (for
$t>s$)
\begin{equation*}
\mathcal{C}(t,s)=\exp(-(t-s)A^0)C+\mathcal{O}\left(
\exp[-(t+s)\zeta/2]\right ),
\end{equation*}
and the effects of the second term in the r.h.s of the equation
above disappears in the correlation formula in the limit of
$t\rightarrow\infty$.

For the anharmonic interaction we choose, for ease of computation,
$\mathcal{P}(\phi)_i(s)=\frac{a_4}{4}:\phi_{i-N}^4(s):$, where the
dots mean the Wick order with respect to the Gaussian measure
$\mu_{\mathcal{C}}$.

We will make a perturbative analysis , i.e., we will assume that
the coupling between two sites $J$ as well as the anharmonic
potential coefficient $\lambda$ are small. Hence, up to first
order in $J$ and $\lambda$, after (considerable but
straightforward)  calculations  we have
\begin{eqnarray}
\lefteqn{\left<\phi_u\phi_v\right>=} \nonumber\\
 && \left\{\begin{array}{ll}\frac{1}{2\zeta M}\left [
\mathcal{J}_{v+N,u-N}T_{u-N}-\mathcal{J}_{u,v}T_v\right
]\delta_{u-N,v} \quad ~~~~\mathrm{for}~~
u\in[N+1,\ldots,2N],v\in[1,\ldots,N]~, \label{tpoint} \\
\\
T_u\delta_{u,v} \quad ~~~~~~~~\mathrm{for}
~~u,v\in[N+1,\ldots,2N].
\end{array} \right .
\end{eqnarray}

For simplicity we will restrict the analysis of the energy current
to one-dimensional systems only. From (\ref{flux}) we have
\begin{equation}
\mathcal{F}_{j<}=\sum_{\begin{array}{c}r>j \\
r\in[1,\ldots,N]\end{array}}\mathcal{J}_{j+N,r}\left (
\phi_j-\phi_r\right )\frac{(\phi_{j+N}+\phi_{r+N})}{2},
\end{equation}
where $\left <\mathcal{F}_{j<} \right >$ denotes de energy flow
between site $j$ and the site $r$ (with $r>j$) connected by the
interaction $\mathcal{J}$. Using the results describe in
(\ref{tpoint}) above we obtain
\begin{equation}
\left <\mathcal{F}_{j<}\right >=\sum_{r>j}\frac{\left
(\mathcal{J}_{j+N,r}\right )^2}{2\zeta M}\left ( T_r-T_j \right ).
\end{equation}

Let us analyze, in particular, the case of next-neighbor
interaction only. In such a case
\begin{equation}
\mathcal{F}_{j\rightarrow j+1}\equiv \left <\mathcal{F}_{j<}\right
>=\frac{\mathcal{J}_{j+N,j+1}}{2\zeta M}\left (T_{j+1}-T_{j}
\right ).
\end{equation}

The condition $\left <dH_i/dt\right
>=0$, that characterizes the stationary state , together with expressions
(\ref{venergy}-\ref{reser}) and  $\left <R_j(t)\right >=0$ (that
comes from (\ref{reser}) and (\ref{tpoint})) lead to

\begin{equation}
\mathcal{F}_{1\rightarrow 2}=\mathcal{F}_{2\rightarrow
3}=\mathcal{F}_{3\rightarrow 4}=\ldots=\mathcal{F}_{N-1\rightarrow
N} .
\end{equation}
I.e., using the notation $J_j\equiv\left (\mathcal{J}_{j+N,j+1}
\right )^2/2\zeta M$,
\begin{equation}
J_1\left( T_2-T_1\right )=J_2\left( T_3-T_2\right )=J_3\left(
T_4-T_3\right )=\ldots=J_{N-1}\left( T_{N}-T_{N-1}\right
).\label{temp}
\end{equation}

It is easy to see that given the temperatures at the boundaries
$T_1$ and $T_N$, and nonvanishing $J_1,J_2,\dots,J_{N-1}$, there
exists an unique solution $T_2,T_3,\ldots,T_{N-1}$ for the linear
system of equations above (\ref{temp}). Namely, we obtain
\begin{equation}
T_k=T_1+\left
(\frac{1}{J_1}+\frac{1}{J_2}+\ldots+\frac{1}{J_{N-1}}\right
)^{-1}\times\left
(\frac{1}{J_1}+\frac{1}{J_2}+\ldots+\frac{1}{J_{k-1}}\right )\left
(T_N-T_1\right ),
\end{equation}
that determines the temperature profile in the steady state. Note
that it is a monotonic function, oriented in the ``right'' way:
the hottest temperature is near the hottest bath, and vice-versa.

For the energy current we get
\begin{equation}
J_{1}(T_{2}-T_{1}) = \ldots = J_j\left (T_{j+1}-T_j\right
)=\chi\frac{\left (T_N-T_1\right )}{N-1}, ~~~~\quad
\frac{\chi}{N-1}=\left
(\frac{1}{J_1}+\frac{1}{J_2}+\ldots+\frac{1}{J_{N-1}}\right
)^{-1},
\end{equation}
that is, the Fourier's law still holds. For the simpler case of
the same interaction between two any next-neighbor sites, i.e.
$J_1=J_2=\ldots=J_{N-1}$, we have, for the thermal conductivity,
\begin{equation}
\chi=J_1=\frac{\left (\mathcal{J}_{1+N,2}\right )^2}{2\zeta
M}.\label{chi}
\end{equation}

For comparison, in \cite{BLL} the authors treat the linear
dynamical problem, i.e. (\ref{dynamics}) with $\lambda=0$ and
\begin{eqnarray}
A=\left (
\begin{array}{cc}
0 & -I \\
\Phi & \zeta
\end{array} \right ), \quad\quad & \begin{array}{ll}
\Phi &=\omega^2(-\Delta+\nu^2) \\
 & =\omega^2(-\delta_{r+1,j}-\delta_{r-1,j}+(2+\nu^2)\delta_{r,j}),\end{array}&
\end{eqnarray}
and obtain (in a nonperturbative approach)
\begin{equation}
\chi=\frac{\omega^2}{\zeta}~\frac{1}{~[2+\nu^2+\sqrt{\nu^2(4+\nu^2)}]}.
\end{equation}
In our case (considering the same $J$ of \cite{BLL}), we have
\begin{eqnarray}
A=\left (
\begin{array}{cc}
0 & -I \\
J+\mathcal{M} & \zeta
\end{array} \right ), \quad\quad & \begin{array}{ll}
J &=\omega^2(-\delta_{r+1,j}-\delta_{r-1,j}), \\
\mathcal{M} &= M\delta_{rj}, ~~M =\omega^2(2+\nu^2),\end{array}&
\end{eqnarray}
and so, our formula (\ref{chi}) above becomes
\begin{equation}
\chi =
\frac{(-\omega^2)^2}{2\zeta(2+\nu^2)\omega^2}=\frac{\omega^2}{\zeta(4+2\nu^2)}
.
\end{equation}
Considering that our computation was carried out in a perturbative
approach with small $J$ but $M$ not small (see the comments at the
final section ), i.e., $\omega^2$ small, $\nu$ large, we have in
(29)
\begin{equation*}
\sqrt{\nu^2(4+\nu^2)}\approx\nu^2\left
(1+\frac{1}{2}\frac{4}{\nu^2}\right )=\nu^2+2.
\end{equation*}
That is, our computation, when restricted to the case treated in
\cite{BLL}, leads to the same result.

In short, we have shown (in a perturbative analysis: up to first
order in $\lambda$ and $J$) that the Fourier's law still holds for
the harmonic crystal with self-consistent reservoirs when a small
nonharmonic on site perturbation is introduced in the interaction.

\section{Concluding Remarks}

The approach presented here establishes an integral representation
for the correlation functions, say, a Feynman-Kac type formalism.
That is, in some sense, we map the stochastic problem on a
non-canonical field theory. Such an approach is inspired on
previous works considering the study of the relaxation to
equilibrium of some nonconservative stochastic Langevin systems
\cite{SBFP}, \cite{FOPS}, \cite{PPLA}, \cite{PPRE}, \cite{PPD}.
There, the time decay of the two and four-point functions is
analyzed in detail. A perturbative study is carried out within the
integral formalism in the regimes of low and high temperature. In
the low temperature region, for the system with a weak anharmonic
potential and a bare mass (the coefficient of the local quadratic
term) large enough,  it is proved that the perturbative analysis
is not naive: e.g., the rigorous results described in \cite{FOPS}
show that the complete treatment of the four-point function adds
only small corrections to the behavior obtained by the
perturbative calculations presented in \cite{SBFP}. Using similar
techniques (cluster expansions, etc) we expect to prove the
results about the behavior of the correlations presented here (for
small $\lambda$, nonzero $M$ and $\zeta$, and $T_{j}$ not large).
The perturbative analysis of our system with all the reservoirs at
(different but) high temperature (i.e., with the perturbative
parameter given by $1/T_{j}$ instead of $\lambda$) shall be
possible following procedures similar to those described in
\cite{PPRE} and references there in. 

Another interesting open problem is the behavior of the system in
the limit of the coupling with the interior heat bath taken to
zero: in such a case, as we have mentioned before (compare
\cite{HuPRE} and \cite{AKAnPh} with \cite{LSJSP}), it is not clear
 if the Fourier's law is valid or not.

\vspace{.5cm} {\bf Acknowlegment.}  Work partially supported by
CNPq and CAPES (Brazil).

\end{document}